\documentclass{article}

\usepackage{neurips_2021}

\usepackage[utf8]{inputenc} 
\usepackage[T1]{fontenc}    
\usepackage{hyperref}       
\usepackage{url}            
\usepackage{booktabs}       
\usepackage{amsfonts}       
\usepackage{nicefrac}       
\usepackage{microtype}      
\usepackage{xcolor}         
\usepackage{svg}

\title{Aiding Medical Diagnosis Through the Application of Graph Neural Networks to Functional MRI Scans}

\author{%
  Katharina Z{\"u}hlsdorff\\
  Department of Psychology/Department of Applied Mathematics and Theoretical Physics\\
  University of Cambridge\\
  Cambridge, UK \\
  \texttt{kz294@cam.ac.uk} \\
   \And
   Clayton Rabideau \\
   Syntensor \\
   \texttt{clayton@syntensor.com} \\
}

\begin{document}

\maketitle
\begin{nolinenumbers}

\begin{abstract}
Graph Neural Networks (GNNs) have been shown to be a powerful tool for generating predictions from biological data. Their
application to neuroimaging data such as functional magnetic resonance imaging (fMRI) scans has been limited. However, applying
GNNs to fMRI scans may substantially  improve predictive accuracy and could be used to inform clinical diagnosis in the future.
In this paper, we present a novel approach to representing resting-state fMRI data as a graph containing nodes and edges
without omitting any of the voxels and thus reducing information loss. We compare multiple GNN architectures and show that they
can successfully predict the disease and sex of a person. We hope to provide a basis for future work to exploit the
power of GNNs when applied to brain imaging data.
\end{abstract}

\section{Introduction}

The application of Graph Neural Networks to biological data has received great attention in recent years (Gilmer et al, 2017; Muzio et al, 2020). Their
ability to predict characteristics of agents such as chemicals and other molecules has proven to be successful, as high predictive
properties have been reported. However, thus far, success in this field has mainly been demonstrated on graph-structured biological data of relatively
small size. Only few groups have tried to extrapolate these findings to brains, which due to their comparatively large size provide an additional
layer of complexity (Li et al, 2020; Kim \& Ye, 2020; Filip et al, 2020).

Functional magnetic resonance imaging (herein referred to as fMRI) is a widely used method to study brain activity, either
when it is at rest (resting-state; rs-fMRI) or when performing a task (task-based fMRI). It is frequently used as a research tool
in neuroscientific research and can aid medical diagnosis, and has
contributed immensely to our understanding of the brain. However, fMRI is not yet being used as a diagnostic tool for
immediate use in the clinic due to the time required to interpret the data and its complexity.
Researchers have developed various algorithms for neuroimaging scans and applied these to fMRI data, with the goal of supporting
medical diagnosis by generating quick predictions (Meszlényi et al., 2017; Patel et al., 2016). However,
applying these to fMRI has proven to be more difficult than initially expected. Some reasons for this include the high dimensionality of the data and
large inter-subject variability, as well as noise in the fMRI time series. Thus, there seems to be an
unmet need for technologies that can successfully employ fMRI as a diagnostic tool.

In this paper, we propose a novel approach to learning from and predicting disease from fMRI data. Our method leverages the power of Graph Neural Networks,
as well as the fact that they can be applied to to graph-structured data, which the brain can be represented as. This `Brain Graph' $\mathbf{G} =(\mathbf{X},\mathbf{A})$ where $\mathbf{X} = \left\{ \vec{\mathbf{X}}_1, \vec{\mathbf{X}}_2, \dots, \vec{\mathbf{X}}_n  \right\}$ , where $n$ is the number of voxels in the fMRI image (representing brain activity in a given region) with time steps $t$ and $\vec{\mathbf{X}}_i \in {\mathbb{R}^t}$, and $\mathbf{A} \in \mathbb{R}^{n \times n}$ is an adjacency matrix generated by a bounded Pearson's Correlation Coefficient function described in Eq. 1.

Previous approaches have combined voxels into so-called
regions-of-interest by averaging them to produce a brain graph with fewer nodes. While this lossy approach results in a lower compute cost when calculating correlation coefficients, models constructed downstream from the application of this filter tend to suffer from poor predictive power. Instead, we use a lossless voxel-wise approach
that removes this data-reduction step.
We generate these graphs
from two publicly available resting-state fMRI datasets and test various different GNN architectures to
predict disease, in this particular example Autism or Attention Deficit Hyperactivity Disorder (ADHD). Our results reveal that GNNs are powerful tools for predicting disease from
rs-fMRI data.

\section{Methods}
\subsection{Datasets and preprocessing}
971 and 794 preprocessed rs-fMRI scans from the publicly accessible Autism Brain Imaging Data Exchange (ABIDE) and
ADHD-200 datasets were used (DiMartino, 2014; Bellec, 2017). These datasets contain preprocessed scans of control participants and subjects with
either Autism or Attention Deficit Hyperactivity Disorder (ADHD), respectively. The signal from each voxel was normalised
and voxels with no signal were filtered out.
Each node contained the node features ${x_v}$, which were the coordinates and time series information from one voxel.
The number of voxels per scan that were converted to a graph that would be fed into our network varied between 60,000-80,000.
The Pearson's Correlation Coefficient (PCC) was calculated between
all of the voxels' time series
using the Fast-GPU-PCC algorithm (Eslami \& Saeed, 2018).
The edge features $e_{vw}$ for the graph were represented by the PCC, which determines the linear
association between two variables x and y:

\begin{center}
\begin{equation}
p_{xy} \ = \ \frac{\sum\limits_{i=1}^T (x_i - \bar{x})(y_i - \bar{y})}{\sqrt{(x_i - \bar{x})^2}\sqrt{(y_i - \bar{y})^2}}
\end{equation}
\end{center}

This was done on an instance with
16 vCPUs and 1 NVIDIA Tesla M60 GPU. The number of all the possible combinations
equals N(N-1)/2, where N is the number of voxels. For a graph containing 60,000 voxels and thus the same number of nodes, this equals 1,799,970,000
correlation values. In order to limit the total number of edges in the graph and only keep the connections of high
importance, only edges with correlation values of higher than 0.9 were included. In the final graphs, there were
approximately 100 edges per node. The edges and nodes were loaded as a Pytorch Geometric object and used for subsequent
experiments.

\begin{figure}[ht!]
    \centering
    \includegraphics[width=1.0\textwidth]{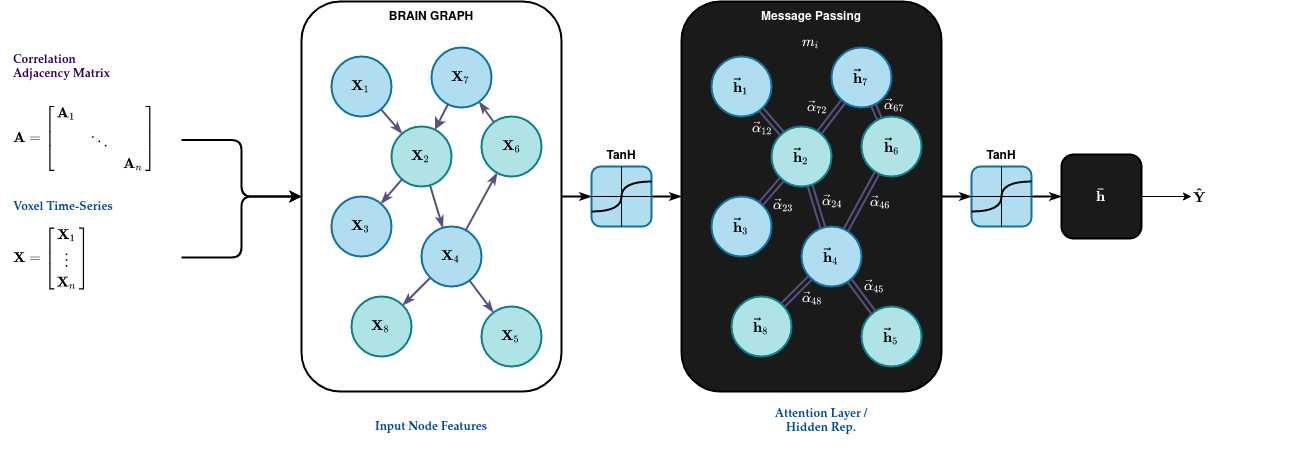}
    \caption{
    A high-level overview of our message-passing network (MPN) based model. Here $m_i$ is one of three types of MPNs: a GCRN, GCN, or GAT. Node hidden states $\mathbf{H} = \left\{ \vec{\mathbf{h}}_1, \vec{\mathbf{h}}_2, \dots, \vec{\mathbf{h}}_n  \right\}$ are averaged post-message-passing for down-stream binary classification.
    }
    \label{fig:GAT}
\end{figure}

\subsection{Comparison of neural networks}
For the purposes of this paper, we tested performance on graph-level, rather than node-level, predictions.
Three different state-of-the-art GNN architectures were tested on these data, including Graph Convolutional Networks (GCN),
Graph Attention Networks (GAT) and Graph Convolutional Recurrent Networks (GCRN) (Kipf \& Welling, 2016; Velickovic et al, 2017; Seo et al, 2016).
Performance of these algorithms was compared to
a simple feedforward network that contained two two linear layers.
All network architectures were implemented in Pytorch Geometric and Pytorch Geometric Temporal (Fey \& Lenssen, 2019; Rozemberczki et al, 2021).
The optimal hyperparameters, which included the learning rate, batch size and number of units in a layer,
were determined using a grid search. The data were separated into train,
validation and test sets (80\%, 10\% and 10\% of each dataset, respectively). The reported values are based on performance achieved on the test set.
We tested performance on both a binary classification task of
the medical condition (i.e. ADHD or Autism), as well as sex (female or male). Performance measures for comparison included the accuracy, F1 and AUROC scores.

\section{Results}

Table 1 summarises the performance of the different architectures tested on the two datasets and on two classification
tasks, which were the medical condition and the sex of the subject.
The network that included GAT layers performed the best out of all of the architectures tested,
with the classification accuracy reaching 0.74 when classifying
ADHD, and 0.87 when classifying sex on the ABIDE dataset. Interestingly, it could predict
Autism better than ADHD. The network including GCN layers performed second best, with accuracies
higher than the GCRN and FFN. The simple FFN had the lowest accuracy
out of all the networks tested, which confirms the predictive abilities of GNNs on these data.

\bigskip

\begin{nolinenumbers}
\begin{center}
\begin{tabular}{c || c | c || c | c}

 Model & Autism & Sex (ABIDE) & ADHD & Sex (ADHD-200)\\ [0.5ex]
 \hline\hline
 FFN & 0.51 & 0.61 & 0.58 & 0.51 \\ [0.5ex]
 \hline
 GCRN & 0.56 & 0.52 & 0.55 & 0.59 \\ [0.5ex]
 \hline
 GCN & 0.63 & 0.83 & 0.69 & 0.81 \\ [0.5ex]
 \hline
 GAT & 0.65 & 0.87 & 0.74 & 0.85 \\ [0.5ex]

\end{tabular}
\end{center}
Table 1: Performance of the different neural network architectures tested using accuracy as a measure.
The algorithms used in these experiments were: a Feedforward Network (FFN), a Graph Convolutional Recurrent Network (GCRN), a Graph Convolutional
Network (GCN) and a Graph Attention Network (GAT). Performance
was tested on two datasets, ABIDE and ADHD-200. They were trained to classify disease and sex.
\end{nolinenumbers}

\section{Conclusion}

In this paper, we demonstrate that resting-state fMRI data can be represented as a graph containing all voxels
of an fMRI scan as well as their respective time series and  does not require the removal or averaging of
voxels, which can lead to loss of information and may not always provide
a fully accurate representation of brain activity. Even though this brain graph contains thousands of nodes and edges,
we prove that it is possible to analyse it using GNNs and make predictions. To our knowledge,
this is a novel approach to analysing rs-fMRI scans. We find that compared to
FFNs, GCRNs and GCNs, GATs had the highest performance on classification tasks. GATs classified scans from patients with ADHD
versus controls with an accuracy of 74\% compared to GCNs (65\%), indicating that they are better suited for this particular task.

We anticipate that the dataset described here will be useful for the characterization of future novel graph methods. We also foresee that the reported accuracy can
be further improved and has potential for clinical application.
We showed that in both disease and sex classification tasks, GATs show state-of-the-art-performance
compared to previously reported studies (Filip et al, 2020; Kim \& Ye, 2020). Additionally,
our results demonstrate that using a voxel-wise approach, rather than a region-of-interest approach, is advantageous
when interpreting fMRI data. We hope that our findings will encourage research groups to make use of the
power of GNNs for analysing brain scans and will adopt a voxel-wise approach in their research.

\begin{nolinenumbers}

\section*{Broader Impact}
As discussed previously, this model has potential as an enabling technology for medical diagnosis and clinical decision making. 

This dataset and its associated construction method are also useful as tools for benchmarking Graph-ML models because brains are a highly complex domain for which there is a lack of pre-constructed datasets. 

It is important to ensure patient confidentiality and data protection if this tool were to be implemented clinically. It is also vital that the patient is aware of
any findings that could arise, is prepared in case a diagnosis is made, and that the correct
support and guidance is provided afterwards. As long as the mentioned points are controlled,
we believe that such tools can be highly beneficial.

\section*{Acknowledgements}
Authors of this project have been supported by the Institute for Neuroscience at the University of Cambridge, UK,
The Alan Turing Institute, London, UK and Syntensor, Inc.

\section*{References}

{
\small
\begin{enumerate}
    \setlength{\leftmargin}{0pt}
    \item Bellec, P., Chu, C., Chouinard-Decorte, F., Benhajali, Y., Margulies, D. S. \& Craddock, R. C. (2017). The Neuro Bureau ADHD-200 Preprocessed repository. NeuroImage, 144, 275–286. https://doi.org/10.1016/j.neuroimage.2016.06.034
    
    \item Di Martino, A., Yan, C. G., Li, Q., Denio, E., Castellanos, F. X., Alaerts, K., Anderson, J. S., Assaf, M., Bookheimer, S. Y., Dapretto, M., Deen, B., Delmonte, S., Dinstein, I., Ertl-Wagner, B., Fair, D. A., Gallagher, L., Kennedy, D. P., Keown, C. L., Keysers, C. \& Milham, M. P. (2014). The autism brain imaging data exchange: Towards a large-scale evaluation of the intrinsic brain architecture in autism. Molecular Psychiatry, 19(6). https://doi.org/10.1038/mp.2013.78
    
    \item Eslami, T. \& Saeed, F. (2018). Fast-GPU-PCC: A GPU-based technique to compute pairwise pearson’s correlation coefficients for time series data—fMRI study. High-Throughput, 7(2). https://doi.org/10.3390/ht7020011
    
    \item Fey, M. \& Lenssen, J. E. (2019). Fast Graph Representation Learning with PyTorch Geometric. ICLR Workshop on Representation Learning on Graphs and Manifolds. https://arxiv.org/abs/1903.02428v3

    \item Filip, A. C., Azevedo, T., Passamonti, L., Toschi, N. \& Lio, P. (2020). A novel Graph Attention Network Architecture for modeling multimodal brain connectivity. Proceedings of the Annual International Conference of the IEEE Engineering in Medicine and Biology Society, EMBS, 2020-July, 1071–1074. https://doi.org/10.1109/EMBC44109.2020.9176613

    \item Gilmer, J., Schoenholz, S. S., Riley, P. F., Vinyals, O. \& Dahl, G. E. (2017). Neural Message Passing for Quantum Chemistry. https://arxiv.org/abs/1704.01212

    \item Hallquist, M. N. \& Hillary, F. G. (2019). Graph theory approaches to functional network organization in brain disorders: A critique for a brave new small-world. Network Neuroscience, 3(1), 1–26. https://doi.org/10.1162/netn\_a\_00054

    \item Kim, B.-H. \& Ye, J. C. (2020). Understanding Graph Isomorphism Network for rs-fMRI Functional Connectivity Analysis. http://arxiv.org/abs/2001.03690

    \item Kipf, T. N. \& Welling, M. (2016). Semi-Supervised Classification with Graph Convolutional Networks. 5th International Conference on Learning Representations, ICLR 2017 - Conference Track Proceedings. https://arxiv.org/abs/1609.02907v4

    \item Li, X., Zhou, Y., Dvornek, N., Zhang, M., Gao, S., Zhuang, J., Scheinost, D., Staib, L. H., Ventola, P. \& Duncan, J. S. (2021). BrainGNN: Interpretable Brain Graph Neural Network for fMRI Analysis. Medical Image Analysis, 102233. https://doi.org/10.1016/J.MEDIA.2021.102233

    \item Meszlényi, R. J., Buza, K. \& Vidnyánszky, Z. (2017). Resting State fMRI Functional Connectivity-Based Classification Using a Convolutional Neural Network Architecture. Frontiers in Neuroinformatics, 0, 61. https://doi.org/10.3389/FNINF.2017.00061

    \item Muzio, G., O’Bray, L. \& Borgwardt, K. (2020). Biological network analysis with deep learning. Briefings in Bioinformatics, 22(2), 1515–1530. https://doi.org/10.1093/bib/bbaa257

    \item Patel, P., Aggarwal, P. \& Gupta, A. (2016). Classification of schizophrenia versus normal subjects using deep learning. ACM International Conference Proceeding Series. https://doi.org/10.1145/3009977.3010050

    \item Rozemberczki, B., Scherer, P., He, Y., Panagopoulos, G., Riedel, A., Astefanoaei, M., Kiss, O., Beres, F., López, G., Collignon, N. \& Sarkar, R. (2021). PyTorch Geometric Temporal: Spatiotemporal Signal Processing with Neural Machine Learning Models. https://arxiv.org/abs/2104.07788v3

    \item Seo, Y., Defferrard, M., Vandergheynst, P. \& Bresson, X. (2016). Structured Sequence Modeling with Graph Convolutional Recurrent Networks. Lecture Notes in Computer Science (Including Subseries Lecture Notes in Artificial Intelligence and Lecture Notes in Bioinformatics), 11301 LNCS, 362–373. https://arxiv.org/abs/1612.07659v1

    \item Veličković, P., Casanova, A., Liò, P., Cucurull, G., Romero, A. \& Bengio, Y. (2018, October 30). Graph attention networks. 6th International Conference on Learning Representations, ICLR 2018 - Conference Track Proceedings. https://arxiv.org/abs/1710.10903v3

\end{enumerate}
}

\end{nolinenumbers}

\end{nolinenumbers}

\end{document}